\documentclass{llncs}

\usepackage{latexsym}
\usepackage{amssymb}
\usepackage{pstricks}

\date{}

  {\begin{center}\begin{minipage}{#1}\hrule\medskip}
  {\vspace{-1ex}\hrule \end{minipage}\end{center}}

\def\lshfit#1#2{\kern-#1 #2\kern#1}

\def\fillsquare{\kern2pt\raise0.25pt
  \hbox{$\vcenter{\hrule height0pt \hbox{\vrule width5pt height5pt} \hrule height0pt}$}
} 

\def\ATS{{\cal A\kern-1ptT\kern-2ptS}}

\begin{document}

\title{%
A Programmer-Centric Approach to\break Program Verification in ATS%
\thanks{~~This work is partially supported by NSF grants No. CCF-0702665.}
} 
\author{Zhiqiang Ren\inst{1} \and Hongwei Xi\inst{1}}
\institute{Boston University}

\maketitle 

\begin{abstract}

Formal specification is widely employed in the construction of high-quality
software. However, there is often a huge gap between formal specification
and actual implementation. While there is already a vast body of work on
software testing and verification, the task to ensure that an
implementation indeed meets its specification is still undeniably of great
difficulty.  ATS is a programming language equipped with a highly
expressive type system that allows the programmer to specify and implement
and then verify within the language itself that an implementation meets its
specification.  In this paper, we present largely through examples a
programmer-centric style of program verification that puts emphasis on
requesting the programmer to explain in a literate fashion why his or her
code works. This is a solid step in the pursuit of software construction
that is verifiably correct according to specification.

\end{abstract}


\setcounter{page}{1}

\baselineskip=11.875pt

\section{Introduction}\label{section:introduction}
In order to be precise in building software systems, we need to specify
what such a system is expected to accomplish. In the current day and age,
software specification, which we use in a rather loose sense, is often done
in forms of varying degree of formalism, ranging from verbal discussions to
pencil/paper drawings to various diagrams in modeling languages such as
UML~\cite{Rayner2005OMG} to formal specifications in specification
languages such as Z~\cite{z:spiv92}, etc. Often the main purpose of
software specification is to establish a mutual understanding among a team
of developers.  After the specification for a software system is done,
either formally or informally, we need to implement the specification in a
programming language.  In general, it is exceedingly difficult to be
reasonably certain whether an implementation actually meets its
specification. Even if the implementation coheres well with its
specification initially, it nearly inevitably diverges from the
specification as the software system evolves. The dreadful consequences of
such a divergence are all too familiar; the specification becomes less and
less reliable for understanding the behavior of the software system while
the implementation gradually turns into its own specification; for the
developers, it becomes increasingly difficult and risky to maintain and
extend the software system; for the users, it requires extra amount of
time and effort to learn and use the software system.

The design of ATS~\cite{ATS-types03,ATSweb} was largely inspired by
Martin-L{\"o}f's constructive type theory~\cite{martinlof85}, which was
originally developed for the purpose of establishing a foundation for
mathematics.  Within ATS, there are a static component (statics) and a
dynamic component (dynamics). Intuitively, the statics and dynamics are
each for handling types and programs, respectively. In particular,
specification is done in the statics and implementation in the
dynamics. Given a specification, how can we effectively ensure that an
implementation of the specification indeed implements according to the
specification?  We expect that the programmer who does the implementation
also constructs a proof in the theorem-proving subsystem of ATS to
demonstrate it.  This is a style of program verification that puts its
emphasis on requesting the programmer to explain in a literate fashion why
his or her code works, and thus we refer to it as a programmer-centric
approach to program verification. The primary contribution of the paper
lies in our effort identifying such a style of program verification as well
as putting it into practice.

In theorem-proving systems such as Coq~\cite{Coq} and
NuPrl~\cite{CONSTABLE86}, a specification is encoded as a type; if a proof
inhabiting the type is made available, then a program guaranteed to meet
the specification can be automatically extracted out of the
proof. Alternatively, formal annotations can be developed for a program
logic such as Hoare logic~\cite{HOARE69} or separation
logic~\cite{SeparationLogic}; a program interspersed with annotations for a
chosen property can be passed to a tool that generates proof obligations
according to the underlying program logic, and the generated proof
obligations can often be discharged through automated theorem-proving. For
instance, KML~\cite{Tushkanova2009Specifying} is a system of this kind for
program verification. The approach to program verification in ATS somewhat
lies in between the two aforementioned ones: It cohesively combines
programming with theorem-proving.

We organize the rest of the paper as follows.  In
Section~\ref{section:ATSoverview}, we give a brief overview of ATS.  We
then present in Section~\ref{section:PVoverview} a typical style of program
verification in ATS that combines programming with theorem-proving.  In
Section~\ref{section:PCV}, we employ some examples to illustrate that ATS
is well-equipped with features to support program verification that is both
flexible and effective for practical use.  Last, we mention some related
work in Section~\ref{section:rwconc} and then conclude.

\def\saddr{\it addr}
\def\sbool{\it bool}
\def\sint{\it int}
\def\schar{\it char}
\def\sfloat{\it float}
\def\snat{\it nat}
\def\sprop{\it prop}
\def\stype{\it type}
\def\sview{\it view}
\def\sviewtype{\it viewtype}

\def\sig{{\cal S}}

\def\emptysig{\sig_{\emptyset}}

\def\emptyssub{[]}
\def\ssub{\Theta_S}
\def\emptydsub{[]}
\def\dsub{\Theta_D}

\def\emptysctx{\emptyset}
\def\emptydctx{\emptyset}

\def\basesort{b}

\def\simp{\rightarrow}
\def\Simp{\Rightarrow}

\def\sconst{\mbox{\it sc}}
\def\dconst{\mbox{\it dc}}

\def\app#1#2{{\bf app}(#1,#2)}
\def\lam#1#2{\mbox{\bf lam}\;#1.#2}

\def\Band{\land}
\def\Bimp{\supset}

\def\timp{\rightarrow}
\def\tint{{\bf int}}
\def\tbool{{\bf bool}}
\def\tptr{{\bf ptr}}
\def\tvoid{{\bf void}}
\def\pf{{\it pf\/}}

\def\pADD{{\bf ADD}}

\begin{figure}
\[\begin{array}{lrcl}
\mbox{sorts} & \sigma & ::= & \basesort \mid\sigma_1\simp\sigma_2 \\
\mbox{static terms} & s & ::= & a \mid \sconst[s_1,\ldots,s_n] \mid \lambda a:\sigma.s \mid s_1(s_2) \\
\mbox{static var. ctx.} & \Sigma & ::= & \emptysctx \mid \Sigma, a:\sigma \\
\mbox{dyn. terms} & d & ::= & x \mid \dconst(d_1,\ldots,d_n) \mid \lam{x}{d} \mid \app{d_1}{d_2} \mid \ldots \\
\mbox{dyn. var. ctx.} & \Delta & ::= & \emptydctx \mid \Delta, x:s \\[6pt]
\end{array}\]
\caption{Some formal syntax for statics and dynamics of ATS}
\label{figure:ATSsyntax}
\end{figure}
\section{Overview of ATS}\label{section:ATSoverview}
We give some formal syntax of ATS in Figure~\ref{figure:ATSsyntax}.  The
language ATS has a static component (statics) and a dynamic component
(dynamics). The statics includes types, props and type indexes while the
dynamics includes programs and proofs. The statics itself is a simply typed
language and a type in it is referred as a {\em sort}. For instance, we
have the following base sorts in ATS: $\saddr$, $\sbool$, $\sint$,
$\sprop$, $\stype$, etc; we use $L$, $B$ and $I$ for static addresses,
booleans and integers of the sorts $\saddr$, $\sbool$ and $\sint$,
respectively; we use $T$ for static terms of the sort $\stype$, which are
types assigned to programs; we use $P$ for static terms of the sort
$\sprop$, which are props assigned to proofs.

Types and props may depend on one or more type indexes of static sorts. A
special case of such indexed types is singleton types, which are each a
type for only one specific value. For instance, $\tbool(B)$ is a singleton
type for the boolean value equal to $B$, and $\tint(I)$ is a singleton type
for the integer equal to $I$, and $\tptr(L)$ is a singleton type for the
pointer that points to the address (or location) $L$. Also, we can quantify
over type index variables universally and existentially to form quantified
types and props.

We use proving-types of the form $(P\mid T)$ for combining proofs with
programs, where $P$ and $T$ stand for a prop and a type, respectively.  One
may think of the proving-type $(P\mid T)$ as a refinement of the type $T$
because $P$ often constrains some of the indexes appearing in $T$. For
example, the following type:
\[\begin{array}{c}
(\pADD(m, n, p) \mid \tint(m), \tint(n), \tint(p))
\end{array}\]
is a proving-type of the sort $\stype$ for a tuple of integers $(m, n, p)$
along with a proof of the prop $\pADD(m, n, p)$ which encodes $m + n = p$.
Given a static boolean term $B$ and a type $T$, we can form two special
forms of types: guarded types of the form $B\Bimp∧T$ and asserting types of
the form $B\Band T$. Following is an example involving singleton, guarded
and asserting types:
\[\begin{array}{l}
\forall a:\sint.a \geq 0 \Bimp (\tint(a)\timp\exists a':\sint. (a' < 0 \Band \tint(a'))) \\
\end{array}\]
The meaning of this type should be clear: Each value that can be assigned this
type represents a function from nonnegative integers to negative integers.

\section{Overview of Program Verification in ATS}\label{section:PVoverview}

\def\fib{{\it fib}}
\def\fibc{{\it fibc}}
\def\fibats{{\it fibats}}
\def\FIB{{\bf FIB}}
\def\FIBzro{{\it FIB0}}
\def\FIBone{{\it FIB1}}
\def\FIBtwo{{\it FIB2}}

We now use a simple example to illustrate the idea of programming with
theorem proving. Suppose we want to compute Fibonacci numbers, which are
defined inductively as follows:
\[\begin{array}{ccccc}
\fib(0) = 0 &\kern12pt& \fib(1) = 1 &\kern12pt& \fib(n+2) = \fib(n) + \fib(n+1)~~\mbox{for $n>=0$} \\
\end{array}\]
A direct implementation of the function {\it fib} in ATS can be done as
follows:
\begin{verbatim}
fun fib (n:int): int =
  if n = 0 then 0 else (if n = 1 then 1 else fib (n-1) + fib (n-2))
// end of [fib]
\end{verbatim}
where the syntax is ML-like. This is a terribly impractical implementation
of exponential time-complexity.  In C, we can give an implementation as
follows that is of O(n) time-complexity:
\begin{verbatim}
int fibc (int n) {
  int tmp, f0 = 0, f1 = 1 ;
  while (n-- > 0) { tmp = f1 ; f1 = f0 + f1 ; f0 = tmp ; } ;
  return f0 ;
} // end of [fibc]
\end{verbatim}
There is obviously a logic gap between the definition of $\fib$ and its
implementation $\fibc$ in C.\footnote{We do not address the issue of
  possible arithmetic overflow here.} In ATS, we can give an implementation
of $\fib$ that completely bridges this gap. First, we need a way to encode
the definition of $\fib$ into ATS, which is fulfilled by the declaration of
the following dataprop:
\begin{verbatim}
dataprop FIB (int, int) =
  | FIB0 (0, 0) | FIB1 (1, 1)
  | {n:nat} {r0,r1:int}
    FIB2 (n+2, r0+r1) of (FIB (n, r0), FIB (n+1, r1))
// end of [FIB]
\end{verbatim}
where the concrete syntax \verb`{...}` is for universal quantification in
ATS. This declaration introduces a type (or more precisely, a type
constructor) $\FIB$ for proofs. Such a type is referred to as a prop (or
prop-type) in ATS. Intuitively, if a proof can be assigned the type
$\FIB(n,r)$ for some integers $n$ and $r$, then $\fib(n)$ equals $r$.  In
other words, $\FIB(n,r)$ encodes the relation $\fib(n)=r$. There are three
constructors $\FIBzro$, $\FIBone$ and $\FIBtwo$ associated with $\FIB$,
which are given the following types corresponding to the three equations in
the definition of $\fib$:
\[\begin{array}{rcl}
\FIBzro & : & () \timp \FIB (0, 0) \\
\FIBone & : & () \timp \FIB (1, 1) \\
\FIBtwo & : & \forall n:nat.\forall r_0:int.\forall r_1:int. \\
        &   & \kern12pt (\FIB (n, r_0), \FIB (n, r_1) \timp \FIB (n+2, r_0+r_1) \\
\end{array}\]
For instance, $\FIBtwo(\FIBzro(), \FIBone())$ is a term of the type
$\FIB(2,1)$, attesting to $\fib(2)=1$.
\begin{figure}
\begin{verbatim}
//
// the syntax [...] is for existential quantification
//
fun fibats {n:nat} (n: int n)
  : [r:int] (FIB (n, r) | int r) = let
  fun loop
    {n,i:nat | i <= n} {r0,r1:int} (
    pf0: FIB (i, r0), pf1: FIB (i+1, r1)
  | r0: int (r0), r1: int (r1), ni: int(n-i)
  ) : [r:int] (FIB (n, r) | int (r)) =
  if ni > 0 then
    loop {n,i+1} (pf1, FIB2 (pf0, pf1) | r1, r0+r1, ni-1)
  else (pf0 | r0)
in
  loop (FIB0(), FIB1() | 0, 1, n)
end // end of [fibats]
\end{verbatim}
\caption{A verified implementation of $\fib$ in ATS}
\label{figure:fibats}
\end{figure}
In Figure~\ref{figure:fibats}, the implemented function $\fibats$ is
assigned the following type:
\[\begin{array}{rcl}
\fibats &:& \forall n:nat.~~\tint(n)\timp\exists r:int.(\FIB(n,r)\mid\tint(r))
\end{array}\]
where $\mid$ is just a separator (like a comma) for separating a proof from
a value.  For each integer $I$, $\tint(I)$ is a singleton type for the only
integer whose value is $I$. When $\fibats$ is applied to an integer of
value $n$, it returns a pair consisting of a proof and an integer value $r$
such that the proof, which is of the type $\FIB(n,r)$, asserts $\fib(n)=r$.
Therefore, $\fibats$ is a verified implementation of $\fib$. Note that the
{\it loop} function in Figure~\ref{figure:fibats} directly corresponds to
the while-loop in the body of $\fibc$.  Also, we emphasize that proofs are
completely erased after typechecking. In particular, there is no proof
construction at run-time.

\def\inil{{\it nil}}
\def\icons{{\it cons}}
\def\silist{{\it ilist}}
\def\tintlist{{\it intlist}}
\section{Programmer-Centric Verification}\label{section:PCV}
By programmer-centric verification, we mean a verification approach that
puts the programmer at the center of the verification process.  The
programmer is expected to explain in a literate fashion why his or her
implementation meets a given specification. The programmer may rely on
external knowledge when doing verification, but such knowledge should be
expressed in a format that is accessible to other programmers.  We will
employ some examples in this section to elaborate on programmer-centric
verification.

\begin{figure}[thp]
\begin{verbatim}
//
// list(a, n) is the type for a list of length n
// in which each element is of the type T.
//
fun{a:type} insort {n:nat}
  (xs: list (a, n), lte: (a, a) -> bool): list (a, n) = let
  fun ins {n:nat}
    (x: a, xs: list (a, n), lte: (a, a) -> bool): list (a, n+1) =
    case xs of
    | list_cons (x1, xs1) =>
        if lte (x, x1) then
          list_cons (x, xs) else list_cons (x1, ins (x, xs1, lte))
        // end of [if]
    | list_nil () => list_cons (x, list_nil ())
  // end of [ins]
in
  case xs of
  | list_cons (x, xs1) => ins (x, insort (xs1, lte), lte)
  | list_nil () => list_nil ()
end // end of [insort]
\end{verbatim}
\caption{A standard implementation of insertion sort}
\label{figure:standard_insertion_sort}
\end{figure}
\begin{figure}
\begin{verbatim}
fun{a:type} insort
  {xs:ilist} (xs: glist (a, xs), lte: lte(a))
  : [ys:ilist] (SORT (xs, ys) | glist (a, ys)) = let
  fun ins {x:int} {ys1:ilist} (
    pford: ORD (ys1) |
    x: E (a, x), ys1: glist (a, ys1), lte: lte(a)
  ) : [ys2:ilist] (SORT (cons (x, ys1), ys2) | glist (a, ys2)) =
    case ys1 of
    | glist_cons (y1, ys10) =>
        if lte (x, y1) then let
          prval pford = ORD_ins {x} (pford)
          prval pfperm = PERM_refl ()
          prval pfsrt = ORDPERM2SORT (pford, pfperm)
        in
          (pfsrt | cons (x, ys1))
        end else let
          prval pford1 = ORD_tail (pford)
          val (pfsrt1 | ys20) = ins (pford1 | x, ys10, lte)
          prval pfsrt2 = SORT_ins {x} (pford, pfsrt1)
        in
          (pfsrt2 | cons (y1, ys20))
        end // end of [if]
    | glist_nil () => (SORT_sing () | cons (x, nil ()))
  // end of [ins]
in
  case xs of
  | glist_cons (x, xs1) => let
      val (pfsrt1 | ys1) = insort (xs1, lte)
      prval pford1 = SORT2ORD (pfsrt1)
      prval pfperm1 = SORT2PERM (pfsrt1)
      prval pfperm1_cons = PERM_cons (pfperm1)
      val (pfsrt2 | ys2) = ins (pford1 | x, ys1, lte)
      prval pford2 = SORT2ORD (pfsrt2)
      prval pfperm2 = SORT2PERM (pfsrt2)
      prval pfperm3 = PERM_tran (pfperm1_cons, pfperm2)
      prval pfsrt3 = ORDPERM2SORT (pford2, pfperm3)
    in
      (pfsrt3 | ys2)
    end // end of [intlist_cons]
  | glist_nil () => (SORT_nil () | nil ())
end // end of [insort]
\end{verbatim}
\caption{A verified implementation of insertion sort}
\label{figure:verified_insertion_sort}
\end{figure}
\subsection{Example: Insertion Sort on Generic Lists}
In Figure~\ref{figure:standard_insertion_sort}, we give a standard
implementation of insertion sort written in ATS that takes a generic list
and a comparison function and returns a generic list that is sorted
according to the comparison function. Note that the use of generic lists
clearly indicates our strive for practicality.  In the literature, a
similar presentation would often use integer lists (instead of generic
lists), revealing the difficulty in handling polymorphism and thus
weakening the argument for practical use of verification. We have no such
difficulty. The implementation we present guarantees based on the types
that the output list is of the same length as the input list. We also give
a verified implementation of insertion sort in
Figure~\ref{figure:verified_insertion_sort} that guarantees based on the
types that the output list is a sorted permutation of the input list. The
fact that this verified implementation can be done in such a concise manner
should yield strong support for the underlying verification approach.

Suppose that a programmer did the implementation in
Figure~\ref{figure:standard_insertion_sort}.  Obviously, the programmer did
not do the implementation in a random fashion; he or she did it based on
some kind of (informal) logic reasoning. We will see that ATS provides
programming features such as abstract props and external lemmas for turning
such informal reasoning into formal verification.  In particular, we can turn
the implementation of insertion sort in
Figure~\ref{figure:standard_insertion_sort} into the verified one in
Figure~\ref{figure:verified_insertion_sort} by following a verification
process.

\def\tE{{\bf E}}
\def\tglist{{\bf glist}}
\def\tlte{{\bf lte}}
\def\pORD{{\bf ORD}}
\def\pPERM{{\bf PERM}}
\def\pSORT{{\bf SORT}}

\begin{figure}[thp]
\begin{verbatim}
abstype E (a:type, x:int) // abstract type constructor
datasort ilist = ilist_nil of () | ilist_cons of (int, ilist)
datatype glist (a:type, ilist) =
  | {x:int} {xs:ilist}
    glist_cons (a, cons (x, xs)) of (E (a, x), glist (a, xs))
  | glist_nil (a, nil) of ()
\end{verbatim}
\caption{A generic list type indexed by the names of list elements}
\label{figure:glist}
\end{figure}
In Figure~\ref{figure:glist}, we first introduce an abstract type
constructor $E$. Given a type $T$ and an integer $I$, $\tE(T,I)$ is a {\em
  singleton} type for a value of the type $T$ with an (imaginary) integer
name $I$.  In ATS, the user-defined sorts (datasorts) can be introduced in
a manner similar to the introduction of user-defined types (datatypes) in a
ML-like language. We introduce a datasort $\silist$ for representing
sequences of (static) integers.  We may simply write {\it nil} and {\it
  cons} for {\it ilist\_nil} and {\it ilist\_cons}, respectively, if there
is no potential confusion.  Note that there is no mechanism for defining
recursive functions in the statics, and this is a profound restriction that
give rise to a unique style of verification in ATS. We lastly define a
datatype $\tglist$: Given a list of values of types $\tE(T, I_1),\ldots,
\tE(T, I_n)$, the type $\tglist (T, cons(I_1,\ldots,cons(I_n,nil)))$ can be
assigned to this particular list.  We may also simply write {\it nil} and
{\it cons} for {\it glist\_nil} and {\it glist\_cons}, respectively, if
there is no potential confusion. Please note that $\tglist$ is in the
dynamics while $\silist$ is in the statics.

To verify insertion sort, we first introduce an abstract prop as follows
such that $\pSORT(xs, ys)$ means that $ys$ is a sorted permutation of $xs$:
\begin{verbatim}
absprop SORT (xs:ilist, ys:ilist)
\end{verbatim}
Let $\tlte(a)$ be a shorthand for the following type:
\[\begin{array}{l}
\forall a:type\forall x_1:\sint\forall x_2:\sint.(\tE (a, x_1), \tE (a, x_2))\timp \tbool(x_1\leq x_2)
\end{array}\]
If we can assign the following type to {\it insort}:
\[
\begin{array}{l}
\forall a:type\forall xs:\silist. \\
\kern12pt (\tglist(a, xs), \tlte (a))\timp \exists ys:\silist.(\pSORT(xs, ys)\mid\tglist(a, ys)) \\
\end{array}
\]
then {\it insort} is verified as the type simply states that the output
list is a sorted permutation of the input list.

For the purpose of verification, we also introduce the following two
abstract props:
\begin{verbatim}
absprop ORD (xs:ilist)
absprop PERM (xs:ilist, ys:ilist)
\end{verbatim}
Given $xs$ and $ys$, $\pORD(xs)$ means that $xs$ is ordered according to
the ordering $\leq$ on integers and $\pPERM(xs, ys)$ means that $ys$ is a
permutation of $xs$.

\begin{figure}[thp]
\[\begin{array}{rcl}
{\it SORT2ORD} &~:~& \forall xs:\silist\forall ys:\silist.~\pSORT(xs, ys) \timp \pORD (ys) \\
{\it SORT2PERM} &~:~& \forall xs:\silist\forall ys:\silist.~\pSORT(xs, ys) \timp \pPERM (xs, ys) \\
{\it ORDPERM2SORT} &~:~&
\forall xs:\silist\forall ys:\silist.~\\
                   & &
\kern6pt(\pORD(ys), \pPERM(xs, ys)) \timp \pSORT (xs, ys) \\
{\it SORT\_nil} &~:~& () \timp \pSORT (nil, nil) \\
{\it SORT\_sing} &~:~& \forall x:\sint.~() \timp \pSORT (cons (x, nil), cons (x, nil)) \\
{\it ORD\_tail} &~:~& \forall y:\sint\forall ys:\silist.~\pORD (cons (y, ys)) \timp \pORD (ys) \\
{\it ORD\_ins} &~:~&
\forall x:\sint\forall y:\sint\forall ys:\silist.~x\leq y\Bimp \\
               & &
\kern6pt\pORD(cons(y, ys))\timp \pORD(cons (x, cons (y, ys))) \\
{\it PERM\_refl} &~:~& \forall xs:\silist.~() \timp \pPERM (xs, xs) \\
{\it PERM\_tran} &~:~&
\forall xs:\silist\forall ys:\silist\forall zs:\silist.~\\
                 & &
\kern6pt(\pPERM(xs, ys), \pPERM(ys, zs))\timp \pPERM(xs, zs) \\
{\it PERM\_cons} &~:~&
\forall x:\sint\forall xs_1:\silist\forall xs_2:\silist.~\\
                 & &
\kern6pt\pPERM(xs_1, xs_2)\timp\pPERM (cons (x, xs_1), cons (x, xs_2)) \\
{\it SORT\_ins} &~:~&
\forall x:\sint\forall y:\sint\forall ys_1:\silist\forall ys_2:\silist.~x>y\Bimp\\
                & &
\kern6pt(\pORD (cons (y, ys_1)), \pSORT (cons (x, ys_1), ys_2)) \timp \\
                & &
\kern6pt\pSORT(cons (x, cons (y, ys_1)), cons (y, ys_2)) \\
\end{array}\]
\caption{Some external lemmas needed for verifying insertion sort}
\label{figure:insort_extern_lemmas}
\end{figure}
\begin{figure}[thp]
\begin{itemize}
\item{\it SORT2ORD}: If $ys$ is a sorted version of $xs$, then $ys$ is ordered
\item {\it SORT2PERM}: If $ys$ is a sorted version of $xs$, then $ys$ is a
  permutation of $xs$
\item {\it ORDPERM2SORT}: if $ys$ is ordered and is also a permutation
  of $xs$, then $ys$ is a sorted version of $xs$.
\item {\it SORT\_nil}: The empty list is a sorted version of itself.
\item {\it SORT\_sing}: A singleton list is a sorted version of itself.
\item
{\it ORD\_tail}: If a non-empty list is ordered, then its tail is also ordered.
\item
{\it ORD\_ins}: If $x\leq y$ holds and $cons(y, ys)$ is ordered, then
$cons(x, cons (y, ys))$ is also ordered.
\item {\it PERM\_refl}: Each list is a permutation of itself
\item {\it PERM\_tran}: The permutation relation is transitive.
\item {\it PERM\_cons}: If $xs_2$ is a permutation of $xs_1$, then
$cons (x, xs_2)$ is a permutation of $cons (x, xs_1)$.
\item {\it SORT\_ins}: If $x>y$ holds, $cons(y, ys_1)$ is ordered and
$ys_2$ is a sorted version of $cons(x, ys_1)$, then $cons (y, ys_2)$ is a
sorted version of $cons (x, cons (y, ys_1))$.
\end{itemize}
\caption{Some explanation for the lemmas in Figure~\ref{figure:insort_extern_lemmas}}
\label{figure:explanation_for_insort_extern_lemmas}
\end{figure}
When verifying {\it insort}, we essentially try to justify each step in the
code presented in Figure~\ref{figure:standard_insertion_sort}. This
justification process may introduce various external lemmas. For instance,
the code presented in Figure~\ref{figure:verified_insertion_sort} makes use
of the lemmas listed in Figure~\ref{figure:insort_extern_lemmas}.

In order to prove these lemmas, we need to define $\pSORT$, $\pORD$ and
$\pPERM$ explicitly, and we can indeed do this in the theorem-proving
subsystem of ATS. However, this style of verifying everything from basic
definitions can be too great a burden in practice. Suppose that we try to
construct a mathematical proof and we need to make use of the proposition
in the proof that the standard permutation relation is transitive. It is
unlikely that we provide an explicit proof for this proposition as {\em it
  sounds so evident to us}. To put it from a different angle, if
constructing mathematical proofs required that every single detail be
presented explicitly, then studying mathematics would unlikely to be
feasible. Therefore, we strongly advocate a style of theorem-proving in ATS
that models the way we do mathematics.

The implementation of insertion sort on generic lists in
Figure~\ref{figure:standard_insertion_sort}, which can be obtained from
erasing proofs in Figure~\ref{figure:verified_insertion_sort}, is
guaranteed to be correct if all of the lemmas in
Figure~\ref{figure:insort_extern_lemmas} are true. Some explanation of these
lemmas is given in Figure~\ref{figure:explanation_for_insort_extern_lemmas}.  It
is probably fair to say that these lemmas are all evidently true except
the last one: {\it SORT\_ins}.  If we are unsure whether the lemma {\it
  SORT\_ins} is true or not, we can construct a proof in ATS or elsewhere
to validate it. For instance, we can even give an informal proof as
follows: {
Note that $\pPERM(cons (x, ys_1), ys_2)$ holds as $ys_2$ is a sorted
version of $cons(x, ys_1)$. Hence, $cons (y, ys_2)$ is a permutation of
$cons (x, cons (y, ys_1))$. Since $cons (y, ys_1)$ is ordered, $y$ is a
lower bound for the elements in $ys_1$. Hence, $y$ is a lower bound for
elements in $ys_2$ as $x>y$ holds, and thus, $cons (y, ys_2)$ is ordered.
Therefore, $cons (y, ys_2)$ is a sorted version of 
$cons (x, cons (y, ys_1))$.
} 

What is of crucial importance is that {\it SORT\_ins} is a lemma that is
{\em manually} introduced and can be readily understood by any programmer
with adequate training. This is a direct consequence of programmer-centric
verification in which the programmer explains in a literate fashion why his
or her implementation meets a given specification.

\def\pLB{{\bf LB}}
\def\pUB{{\bf UB}}
\def\pUNION{{\bf UNION4}}
\def\pAPPEND{{\bf APPEND}}
\def\mset#1{|#1|}

\begin{figure}[thp]
\begin{verbatim}
fun{a:type}
qsrt {n:nat}
  (xs: list (a, n), lte: lte a) : list (a, n) =
  case+ xs of
  | list_cons (x, xs) => part (x, xs, lte, list_nil (), list_nil ())
  | list_nil () => list_nil ()

and part {p:nat} {q,r:nat} (
  x0: a, xs: list (a, p), lte: lte(a), ys: list (a, q), zs: list (a, r)
) : list (a, p+q+r+1) =
  case+ xs of
  | list_cons (x, xs) =>
      if lte (x, x0) then
        part (x0, xs, lte, list_cons (x, ys), zs)
      else
        part (x0, xs, lte, ys, list_cons (x, zs))
      // end of [if]        
  | list_nil () => let
      val ys = qsrt (ys, lte) and zs = qsrt (zs, lte)
    in
      append (ys, list_cons (x0, zs))
    end // end of [list_nil]
\end{verbatim}
\caption{A standard implementation of quicksort}
\label{figure:standard_quicksort}
\end{figure}

\begin{figure}[thp]
\begin{verbatim}
fun{a:type}
qsrt {xs:ilist} (
  xs: glist (a, xs), lte: lte a
) : [ys:ilist] (SORT (xs, ys) | glist (a, ys)) =
  case+ xs of
  | glist_cons (x, xs) => let
      val (pford, pfuni | res) =
        part (UB_nil (), LB_nil () | x, xs, lte, nil (), nil ())
      prval pfperm = UNION4_perm (pfuni)
    in
      (ORDPERM2SORT (pford, pfperm) | res)
    end
  | glist_nil () => (SORT_nil () | nil ())

and part
  {x0:int} {xs:ilist} {ys,zs:ilist} (
  pf1: UB (x0, ys), pf2: LB (x0, zs)
| x0: E (a, x0), xs: glist (a, xs), lte: lte(a)
, ys: glist (a, ys), zs: glist (a, zs)
) : [res:ilist] (
  ORD (res), UNION4 (x0, xs, ys, zs, res) | glist (a, res)
) =
  case+ xs of
  | glist_cons (x, xs) =>
      if lte (x, x0) then let
        prval pf1 = UB_cons (pf1)
        val (pford, pfuni | res) =
          part (pf1, pf2 | x0, xs, lte, cons (x, ys), zs)
        prval pfuni = UNION4_mov1 (pfuni)
      in
        (pford, pfuni | res)
      end else let
        prval pf2 = LB_cons (pf2)
        val (pford, pfuni | res) =
          part (pf1, pf2 | x0, xs, lte, ys, cons (x, zs))
        prval pfuni = UNION4_mov2 (pfuni)
      in
        (pford, pfuni | res)
      end // end of [if]        
  | glist_nil () => let
      val (pfsrt1 | ys) = qsrt (ys, lte)
      val (pfsrt2 | zs) = qsrt (zs, lte)
      val (pfapp | res) = append (ys, cons (x0, zs))
      prval pford1 = SORT2ORD (pfsrt1)
      prval pford2 = SORT2ORD (pfsrt2)
      prval pfperm1 = SORT2PERM (pfsrt1)
      prval pfperm2 = SORT2PERM (pfsrt2)
      prval pf1 = UB_perm (pfperm1, pf1)
      prval pf2 = LB_perm (pfperm2, pf2)
      prval pford = APPEND_ord (pf1, pf2, pford1, pford2, pfapp)
      prval pfuni = APPEND_union4 (pfperm1, pfperm2, pfapp)
    in
      (pford, pfuni | res)
    end // end of [glist_nil]
// end of [part]
\end{verbatim}
\caption{A verified implementation of quicksort}
\label{figure:verified_quicksort}
\end{figure}

\begin{figure}[thp]
\[\begin{array}{rcl}
{\it LB\_nil} &~:~& \forall x:\sint.~() \timp \pLB (x, nil) \\
{\it UB\_nil} &~:~& \forall x:\sint.~() \timp \pUB (x, nil) \\
{\it LB\_cons} &~:~& \forall x_0:\sint\forall x:\sint\forall xs:\silist.~x_0\leq x\Bimp\\
                  & &
\pLB (x_0, xs) \timp \pLB (x_0, cons (x, xs)) \\
{\it UB\_cons} &~:~& \forall x_0:\sint\forall x:\sint\forall xs:\silist.~x_0\geq x\Bimp\\
                  & &
\pUB (x_0, xs) \timp \pUB (x_0, cons (x, xs)) \\

{\it LB\_perm} &~:~&
\forall x:\sint\forall xs_1:\silist\forall xs_2:\silist.~ \\
               & &
(\pPERM (xs_1, xs_2), \pLB (x, xs_1))\timp \pLB (x, xs_2) \\
{\it UB\_perm} &~:~&
\forall x:\sint\forall xs_1:\silist\forall xs_2:\silist.~ \\
               & &
(\pPERM (xs_1, xs_2), \pUB (x, xs_1))\timp \pUB (x, xs_2) \\

{\it UNION4\_perm} &~:~&
\forall x:\sint\forall xs:\silist\forall res:\silist.~ \\
                   & &
\kern6pt\pUNION (x, xs, nil, nil, res) \timp \pPERM (cons (x, xs), res) \\

{\it UNION4\_mov1}  &~:~&
\forall x_0:\sint\forall x:\sint\forall xs:\silist\forall ys:\silist\forall zs:\silist\forall res:\silist.~\\
                   & &
\kern6pt\pUNION (x0, xs, cons (x, ys), zs, res)\timp\\
                   & &
\kern6pt\pUNION (x0, cons (x, xs), ys, zs, res) \\
{\it UNION4\_mov2}  &~:~&
\forall x_0:\sint\forall x:\sint\forall xs:\silist\forall ys:\silist\forall zs:\silist\forall res:\silist.~\\
                   & &
\kern6pt\pUNION (x0, xs, ys, cons (x, zs), res)\timp\\
                   & &
\kern6pt\pUNION (x0, cons (x, xs), ys, zs, res) \\
{\it APPEND\_ord} &~:~&
\forall x:\sint\forall ys:\silist\forall zs:\silist\forall res:\silist.~\\
                  & &
\kern6pt(\pUB (x, ys), \pLB (x, zs), \pORD(ys), \pORD(zs), \\
                  & &
\kern6pt~\pAPPEND (ys, cons (x, zs), res)) \timp \pORD(res) \\
{\it APPEND\_union4} &~:~&
\forall x:\sint\forall ys:\silist\forall ys_1:\silist\forall
zs:\silist\forall zs_1:\silist\forall res:\silist.~\\
                  & &
\kern6pt(\pPERM(ys, ys_1), \pPERM (zs, zs_1), \\
                  & &
\kern6pt~\pAPPEND (ys_1, cons (x, zs_1), res))\timp\\
                  & &
\kern6pt\pUNION(x, nil, ys, zs, res) \\
\end{array}\]
\caption{Some external lemmas needed for verifying quicksort}
\label{figure:quicksort_extern_lemmas}
\end{figure}
\begin{figure}[thp]
\begin{itemize}


\item{\it LB\_perm}: If $x$ is a lower bound for $xs_1$ and
$xs_1$ is a permutation of $xs_2$, then $x$ is also a lower bound for $xs_2$.
\item{\it UB\_perm}: If $x$ is an upper bound for $xs_1$ and
$xs_1$ is a permeation of $xs_2$, then $x$ is also a upper bound for $xs_2$.

\item{\it UNION4\_perm}:
If $\mset{res} = \{x\}\cup\mset{xs}$, then $res$ is a permutation of $cons(x, xs)$.

\item{\it UNION4\_mov1}:
If $\mset{res} = \{x_0\}\cup\mset{xs}\cup\mset{cons(x,ys)}\cup\mset{zs}$,
then $\mset{res} = \{x_0\}\cup\mset{cons(x,xs)}\cup\mset{ys}\cup\mset{zs}$.
\item{\it UNION4\_mov2}:
If $\mset{res} = \{x_0\}\cup\mset{xs}\cup\mset{ys}\cup\mset{cons(x,zs)}$,
then $\mset{res} = \{x_0\}\cup\mset{cons(x,xs)}\cup\mset{ys}\cup\mset{zs}$.

\item{\it APPEND\_ord}: If $x$ is an upper bound for $ys$ and a lower bound
for $zs$, both $ys$ and $zs$ are ordered and $res$ is the concatenation
of $ys$ and $cons(x, zs)$, then $res$ is ordered.
\item{\it APPEND\_union4}:
If $ys_1$ is a permutation of $ys$, $zs_1$ is a permutation of $zs$ and
$res$ is the concatenation of $ys_1$ and $cons (x, zs_1)$, then
$\mset{res}=\{x\}\cup\mset{ys}\cup\mset{zs}$.
\end{itemize}
\caption{Some explanation for the lemmas in Figure~\ref{figure:quicksort_extern_lemmas}}
\label{figure:explanation_for_quicksort_extern_lemmas}
\end{figure}
\subsection{Example: Quicksort on Generic Lists}
We give a standard implementation of quicksort on generic lists in
Figure~\ref{figure:standard_quicksort}. The reason that we use lists
instead of arrays is solely for simplifying the presentation. As far as
verification is concerned, there is really not much difference between
lists and arrays. Note that we have already made various verification
examples available on-line that involve arrays.

The implementation in Figure~\ref{figure:standard_quicksort} guarantees
based on the types that the output list is of the same length as the input
list. We also give a verified implementation of quicksort in
Figure~\ref{figure:verified_quicksort} that guarantees based on the types
that the output list is a sorted permutation of the input list. The
verified implementation is essentially obtained from the process to explain
why the function {\it qsrt} in Figure~\ref{figure:standard_quicksort}
always returns a list that is the sorted version of the input list.

We now explain that the verified implementation of quicksort can be
trusted. The function {\it append} in the implementation is given the
following type:
\[\begin{array}{l}
\forall a:type.\forall xs_1:\silist\forall xs_2:\silist.~ \\
\kern6pt(\tglist(a, xs_1), \tglist(a, xs_2))\timp \\
\kern6pt\exists res:\silist.(\pAPPEND(xs_1, xs_2, res) \mid \tglist (a, res)) \\
\end{array}\]
where $\pAPPEND$ is an abstract prop. Given lists $xs_1,xs_2$ and $res$,
the intended meaning of $\pAPPEND(xs_1,xs_2,res)$ is obvious: it states
that the concatenation of $xs_1$ and $xs_2$ is $res$. Both $\pLB$ and
$\pUB$ are introduced as abstract props: $\pLB(x, xs)/\pUB(x,xs)$ means
that $x$ is a lower/upper bound for the elements in $xs$. Another
introduced abstract prop is $\pUNION$: Given $x$, $xs$, $ys$, $zs$ and
$res$, $\pUNION(x, xs, ys, zs, res)$ means that the following equation
holds
\[\begin{array}{rcl}
\mset{res}=\{x\}\cup\mset{xs}\cup\mset{ys}\cup\mset{zs} \\
\end{array}\]
where $\mset{\cdot}$ turns an integer list into a multiset.  The external
lemmas in Figure~\ref{figure:verified_quicksort} are listed in
Figure~\ref{figure:quicksort_extern_lemmas} and some explanation are given
in Figure~\ref{figure:explanation_for_quicksort_extern_lemmas} for some of
these lemmas. 

\subsection{Many Other Examples}
There are also a variety of examples available on-line\footnote{ Please see
  \texttt{http://www.ats-lang.org/EXAMPLE/PCPV} } which can further
illustrate a style of programmer-centric verification in ATS that combines
programming with theorem-proving cohesively. In particular, there are
examples involving arrays, heaps, balanced trees, etc.

\section{Related Work and Conclusion}\label{section:rwconc}
Given the vastness of the field of program verification, we can only
mention some closely related work in this section.

In the Coq theorem-proving system~\cite{Dowek93tr}, programs can be
extracted from proofs~\cite{CoqExtraction}.
In~\cite{FilliatreCertification}, the authors specified the orderedness
property as well as the permutation relation on the array structure and
gave verification for three sorting algorithms. However, Coq is primarily
designed for theorem-proving instead of programming, and its use as a
programming language is a bit unwieldy and limited.

Ynot~\cite{Ynot-EIP} is an axiomatic extension of the Coq
proof assistant for specifying and verifying properties of imperative
programs. The programmer can encode a new domain by providing key lemmas in
an ML-like embedded language.  Relying on Coq to do theorem-proving, Ynot
mixes the automated proof generation with manual proof construction,
attempting to relieve the programmer from the heavy burden that would
otherwise be necessary.

In the specification language KML~\cite{Tushkanova2009Specifying}, the
programmer can add annotations such as preconditions, postconditions and
invariants into Java programs, and a tool is provided for generating proof
obligations automatically from the annotated source file, which are to be
discharged by various automatic provers. Permutation can be specified based on
the natural concept of multiset. Often, external assertions need to be
provided along together with the code so as to make the program verifiable
by existing theorem provers.

The work on extended static checking (ESC)~\cite{ESC} also puts emphasis on
employing formal annotations to capture program invariants. These
invariants may be verified through (light-weighted) theorem
proving. ESC/Java~\cite{Fla02} generates verification-conditions based on
annotated Java code and uses an automatic theorem-prover to reason about
the semantics of the programs. It can catch many basic errors such as null
dereferences, array bounds errors, type cast errors, etc.  With more
emphasis on usefulness, soundness is sacrificed in certain cases to reduce
annotation cost or to improve checking speed.

VeriFast~\cite{VeriFast} is another system for verifying program properties
through source code annotation. It supports direct insertion of simple
proof steps into the source code while allowing rich and complex properties
to be specified through inductive datatypes and fixed-point functions.
VeriFast provides a program verifier for C and Java that supports
interactive insertion of annotations into source code.

The paradigm of programming with theorem-proving as is supported in the ATS
programming language system is a novel invention. In particular, this
programming paradigm is fundamentally different from program extraction
(from proofs) as is supported in theorem-proving systems such as Coq.  In
this paper, we have argued for a style of program verification that puts
emphasis on requesting the programmer to formally explain in a literate
fashion why the code he or she implements actually meets its specification.
Though external lemmas introduced during a verification process can be
discharged by formally proving them in ATS, doing so is often expensive in
terms of effort and time.  One possibility is to characterize such lemmas
into different categories and then employ (external) theorem-provers
specialized for a particular category to prove lemmas in that category.
Another possibility we advocate for discharging lemmas is through a
peer-review process, which mimics the practice of (informally) verifying
mathematical proofs. Obviously, the precondition for such an approach is
that the lemmas to be verified can be expressed in a format that is easily
accessible to a (trained) human being. This is where the programmer-centric
verification as is presented in this paper can fit very well.

\bibliographystyle{is-alpha}
\bibliography{xi}

\end{document}